\documentclass{aa}  
\usepackage{amsmath, amssymb, graphicx, natbib}
\bibpunct{(}{)}{;}{a}{}{,}
\usepackage{txfonts}
\begin{document}
   \title{Detectability of neutral interstellar deuterium by a forthcoming SMEX mission IBEX}

   \author{S. Tarnopolski
          \and
          M. Bzowski  \inst{}
          }

   \offprints{M. Bzowski (\email{bzowski@cbk.waw.pl})}

   \institute{Space Research Centre, Polish Academy of Sciences, Bartycka $18$A, $00$-$716$ Warsaw, Poland\\
              \email{slatar@cbk.waw.pl,bzowski@cbk.waw.pl}
         }

   \date{}

 
   \abstract
   {We study the feasibility of detection of neutral interstellar deuterium by the forthcoming NASA SMEX mission IBEX.}
   {Using numerical simulations, we study the absolute density and flux in Earth orbit of neutral interstellar deuterium and check its detectability using IBEX.}
   {Our simulations were performed using the Warsaw 3D time-dependent test-particle model of neutral interstellar gas in the inner heliosphere, which was specially adapted to the case of deuterium, and state-of-the-art models of the ionization field and radiation pressure. The modeling predicted the  density, bulk velocity, and flux of interstellar D at different positions of the Earth during the solar cycle. We paid  particular attention to the time interval in which IBEX observations will be performed.}
   {Using our simulations, we predict a large enhancement of deuterium abundance in Earth orbit with respect to the abundance at the termination shock. The energy of the D atoms at IBEX will be within the energetic sensitivity band of its Lo instrument, apart from during a short time interval between September and November each year. Because of the specific observing geometry of IBEX, there will be one opportunity each year to search for I/S D, when Earth is close to ecliptic longitude $165\degr$, i.e. in March. Assuming that the TS abundance of D is identical as in the Local Cloud, which is equal to $1.56 \times 10^{-5}$, and that the density of H at TS is $0.11$~cm$^{-2}$~s$^{-1}$, we estimate the expected relative flux to be approximately $0.015$ cm$^{-2}$~s$^{-1}$, which corresponds to the local absolute flux of about $0.007$~cm$^{-2}$~s$^{-1}$. The dependence of the expected flux on the phase of solar cycle is relatively weak. The flux scales proportionally to the density of deuterium at the termination shock and depends only weakly on the bulk velocity and temperature of the gas in this region.}
   {}

\titlerunning{Detectability of interstellar D by IBEX}
\authorrunning{Tarnopolski \& Bzowski}

   \maketitle
%

\section{Introduction}

The abundance of deuterium in the Local Interstellar Medium used to be a subject of debate that appears to have been resolved only a couple of years ago \citep{linsky_etal:06a, linsky:07}. Its total value, which challenges the chemical evolution theory of the Galaxy, is approximately 23 ppm; in the gas phase, however, it is only 15.6 ppm. This knowledge is derived primarily from analyses of observations of UV lines of interstellar matter seen in the spectra of nearby stars, i.e. averaged over parsecs and in most cases over different clouds in the LISM. Technology available today makes it feasible to measure the D density in the LIC by direct detection of neutral deuterium in situ in the inner heliosphere either by observations of derivative populations (D$^+$ pickup ions), or by direct detection of the interstellar D atoms themselves. Hence, a reconnaissance of modifications of density and flux distribution of interstellar D inside the heliosphere, and of its abundance with respect to hydrogen, becomes a relevant and urgent task.

The only study of the distribution of interstellar D in the heliosphere that we are aware of, is due to \citet{fahr:79}, who performed the analysis in the approximation of the perfect hot model of the neutral interstellar gas \citep{fahr:78}. \citet{fahr:79} assumed that the normalized radiation pressure, expressed in units of solar gravity force (the so-called $\mu$-units), acting on D, is equal exactly to half of the normalized pressure acting on H. Fahr predicted a gradual increase in the abundance of D with respect to H with decreasing heliocentric distance and a persistence of the overdensity of D in the downwind region throughout the solar cycle. However, because of the lack of relevant data at this time, an important aspect of the problem was neglected.

Interstellar D is subjected to identical ionization processes in the heliosphere as H. The only differences in factors shaping its distribution in the heliosphere appear to be its lower thermal velocity, because of the higher atomic mass, and a different radiation pressure, because both of the mass difference and the isotope effect that shifts its Lyman-$\alpha$ resonance wavelength by $-0.0333$~nm with respect to the center of the self-reversed solar Lyman-$\alpha$ line. It turns out that the latter effect has far-reaching consequences for the distribution of abundances, densities, and fluxes of interstellar D in the heliosphere, especially in the context of its potential detection by a forthcoming NASA SMEX mission IBEX, scheduled for launch in the Summer of 2008  \citep{mccomas_etal:04a, mccomas_etal:05a, mccomas_etal:06}.

We begin with a brief discussion of the factors affecting the kinematics of neutral D atoms in the inner heliosphere and we present the simulations performed. Subsequently, we  present the distribution of density, bulk velocity, flux, and abundance of neutral interstellar D along the Earth orbit and their variations during the solar cycle. We conclude by presenting the conditions under which interstellar D will be detectable by the IBEX mission.

\section{Model and simulations}

\begin{figure}
\centering
\includegraphics[width=8cm]{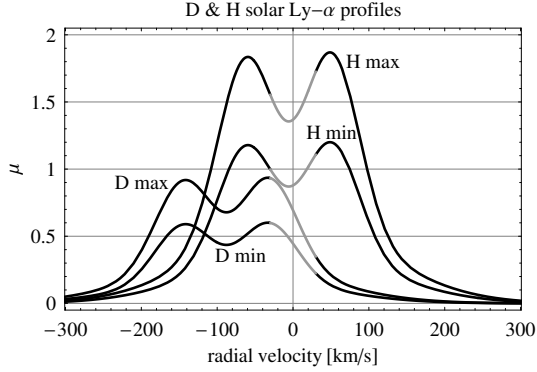}
\caption{Compensation factors of solar gravity due to Lyman-$\alpha$ radiation pressure for H and D atoms for solar minimum and maximum conditions as a function of radial velocity of the atoms. The profiles, scaled in km/s of equivalent Doppler shift and in $\mu$ units, are shown according to the models defined in Eq.(\ref{eq:fifuh}) and Eq.(\ref{eq:fifud}), for the net flux $I_{\mathrm{ tot}}$ equal to $3.53 \times 10^{11}$ cm$^{-2}$~s$^{-1}$ (solar min.) and $5.49 \times 10^{11}$ cm$^{-2}$~s$^{-1}$ (solar max.). Approximate spectral regions relevant for D and H are marked in gray.}
\label{fig:prof}
\end{figure}

Neutral interstellar deuterium atoms in the heliosphere are collisionless and obey the equation of motion that includes the attraction by solar gravity and repulsion by the solar EUV radiation in the Lyman-$\alpha$ line:
\begin{equation}
\label{eq:eom}
d^{2} \vec{r}/{dt^{2}} = -G\, M \left[1 - \mu\left(v_r\left(t\right), I_{\mathrm{tot}}\left(t\right)\right)\right] \vec{r}/|\vec{r}|^3,
\end{equation}
where $\vec{r}\left(t\right)$ is the position vector of the atom with respect to the Sun at time $t$ and $G\,M$ the gravity constant times solar mass. The compensation factor of solar gravity due to radiation pressure $\mu$ depends on the radial velocity of the atom $v_r$ and the line-integrated solar Lyman-$\alpha$ flux $I_{\mathrm{tot}}\left(t\right)$, which was shown for H by \citet{tarnopolski:07} and \citet{tarnopolski_bzowski:07a} using observations of the solar Lyman-$\alpha$ line profiles by \citet{lemaire_etal:02}:
\begin{eqnarray}
\label{eq:fifuh}
\mu \left(v_r, I_{\mathrm{tot}}\right)&=&A\left(1 + B\, I_{\mathrm{tot}}\right) \exp \left(-C v_r^2\right)\\ &\times &\left[1 + D \exp \left(F v_r
- G v_r^2\right) + H \exp \left(-P v_r - Q v_r^2\right)\right] \nonumber
\end{eqnarray}
For deuterium, the compensation factor $\mu_D$ is derived from the formula for hydrogen (\ref{eq:fifuh}) by changing the variable $v_r$ to $v_r - 82.1201$~km/s and scaling the result down by the D/H atomic mass ratio \citep{tarnopolski:07} to obtain:
\begin{eqnarray}
\label{eq:fifud}
	\mu_{D} \left(v_{r}, I_{\mathrm{tot}}\right) & = & a\left(1 + b\, I_{\mathrm{tot}}\right) \left[\exp \left(-c\, v_{r} - d\, v_{r}^{2}\right) +   {}\right. \nonumber \\
	&+ & \left. {} f \exp \left(-g\, v_{r} - h\, v_{r}^{2}\right) + p \exp \left(-q\, v_{r} - r\, v_{r}^{2}\right)\right],
\end{eqnarray}
with the following parameters:

$\begin{array}{lllllll}
a = 4.9469\times 10^{7},  & b = 4.5694\times 10^{-4}, & \\
c = 2.3603\times 10^{-2},  & d = 3.8967\times 10^{-4}, & \\
f = 5.6579\times 10^{-4}, & g = 0.10795,             & h = 3.7205\times 10^{-4}, \\
p = 0.52459,              & q = 6.2923\times 10^{-3},  &  r = 3.8312 \times 10^{-5}.\\
\end{array}$

The normalized radiation pressure (measured in $\mu$-units) acting on D would be just half of the pressure acting on H if the solar Lyman-$\alpha$ line were flat. An isotope effect creates a small difference in the Lyman-$\alpha$ resonance wavelength of the H and D atoms. This difference of $-0.0333$~nm, albeit small, is responsible for the phenomena in the heliospheric gas that we discuss in the paper. Since the solar Lyman-alpha line is not flat, as demonstrated in Fig. \ref{fig:prof}, the spectral fluxes at the resonance Lyman-$\alpha$ wavelengths of H and D atoms are different; when H and D atoms have identical radial velocities, the normalized radiation pressure force that they experience usually differs therfore by a factor other than 0.5. When the radial velocity of a D atom moving towards the Sun is larger than the radial velocity of a H atom, the differences in radiation pressure increase still more.

The range of radial velocities of heliospheric H atoms from the primary and secondary populations \citep{bzowski_etal:97,izmodenov:01} position them within the central self-reversal of the profile: the usual assumption that the radiation pressure does not depend on radial velocity therefore appears reasonable although not perfect \citep{tarnopolski_bzowski:07a}. By contrast, the radial velocities of D fall on the blue slope of the solar line, so the radiation pressure is highly asymmetric with respect to the zero radial velocity. Before perihelion, the atoms are subject to a relatively high radiation pressure -- a little higher than half of the radiation pressure acting on the corresponding H atoms. This ensures that the D atoms accelerate to a higher speed on the pre-perihelion leg of their orbits than the H atoms do. After perihelion, this speed is gradually reduced because the D atoms wander to the wing region of the solar line where the resulting compensation of solar gravity is appreciably reduced with respect to the values at the inbound wing, and the net force is therefore highly attractive. Hence the D atoms are strongly decelerated at the outbound leg of their trajectories and leave the solar vicinity with a lower speed than they had when approaching the Sun, especially during solar minimum. For example, an atom that passes the Sun at a distance of 1~AU traveling at a velocity of approximately 32~km/s and had a velocity of 22~km/s at 100~AU in front of the Sun, has a velocity of about 15~km/s at 100~AU behind the Sun at solar minimum. At solar maximum, it is caught within a bound orbit, where it becomes ionized. For the atoms not trapped in bound orbits, the entrance speed would of course equal the exit speed, if the radiation pressure force had been symmetric with respect to zero radial velocity.

The simulations of densities, bulk velocities, and fluxes of D at 1~AU were performed using the Warsaw test-particle 3D time-dependent model of interstellar gas in the inner heliosphere \citep{rucinski_bzowski:95b, bzowski_etal:97, bzowski_etal:08a}, which had been specially adapted for deuterium \citep{tarnopolski:07}. The classical approach of the hot model \citep{thomas:78} was used, but instead of Keplerian motion the atoms were assumed to follow trajectories modeled using numerical solutions of the equation of motion Eq.(\ref{eq:eom}) in addition to Eq.(\ref{eq:fifud}). The variations in the charge-exchange rate with both latitude and time \citep{bzowski_etal:08a, bzowski:01a}, and the electron impact ionization in a static, spherically symmetric approximation \citep{bzowski_etal:08a}, were all taken into account. The radiation-pressure factor $\mu$ was governed by the net flux $I_{\mathrm{tot}}$ in the solar Lyman-$\alpha$ line, which was taken from the SOLAR 2000 model \citep{tobiska_etal:00c} and approximated by the analytic formula from \citet{bzowski:01a} with parameters from \citet{bzowski_etal:08a}. Details will be shown in a future paper.

The simulations were carried out for the $\sim 11$-year cycle of solar activity 1986--1998 and followed the Earth position during its yearly motion for the 5-th day of each month; the calculation point in June and December of each year therefore coincided with the projection of the inflow direction on the Earth orbit. The upwind direction was assumed to equal the direction calculated by \citet{witte:04}, using in situ observations of interstellar He. The density of interstellar H at the termination shock (TS) was assumed to equal 0.11~cm$^{-3}$ \citep{bzowski_etal:08a} and the abundance of D with respect to H at TS was assumed to be identical to that of the gas phase of the Local Interstellar Cloud$\xi_D = 1.56\times10^{-5}$ \citep{linsky_etal:06a}. The bulk velocity and temperature of the gas at TS were equal to 22~km/s and 12000~K, respectively, close to the average values of these parameters from the primary and secondary populations of interstellar hydrogen from the Moscow MC model adopted by \citet{bzowski_etal:08a} to obtain the H density used in the simulations, and in agreement with values inferred for interstellar H at TS by \citet{costa_etal:99}. 

The problem of filtration of D through the interface has not yet been addressed. \citet{fahr_etal:95} simulated the filtration of interstellar O assuming that it is almost in charge exchange resonance with protons, as is the case for D; they discovered that the balance between the primary and secondary populations at TS is different from the case of hydrogen, which produces net O parameters at TS that are closer to the parameters in the unperturbed gas. \citet{osterbart_fahr:92} showed a breakdown between H populations that originate in various collision hierarchies which, together with the oxygen study, suggest that in the case of D the secondary population may have different parameters from the secondary population of H. These conclusions were supported by simulations of O transmission through the interface by \citet{izmodenov_etal:97a}, who found a transmission factor of O equal to 0.7. To assess the robustness of our results, we therefore completed a parameter study of the 1~AU density of D as a function of the bulk velocity and temperature at TS.

With deuterium simulations, corresponding simulations of hydrogen were also performed.

\section{Results}
\begin{figure}
\centering
\includegraphics[width=8cm]{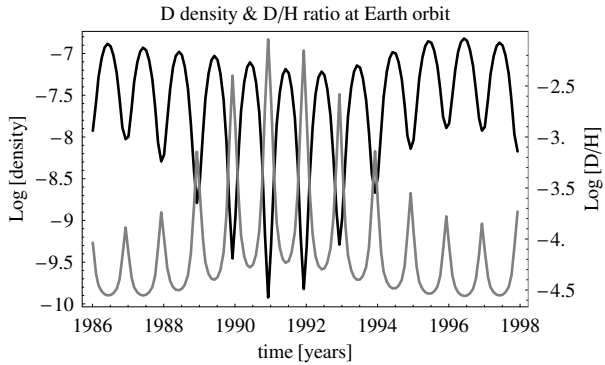} 
\caption{Density of interstellar D along the Earth orbit during the solar cycle (black, left-hand scale) and abundance of D with respect to H (gray, right-hand scale) for the D/H TS abundance $1.56\times 10^{-5}$ and H density 0.11~cm$^{-3}$. }
 \label{fig:dens}
\end{figure}
\begin{figure}
\centering
\includegraphics[width=8cm]{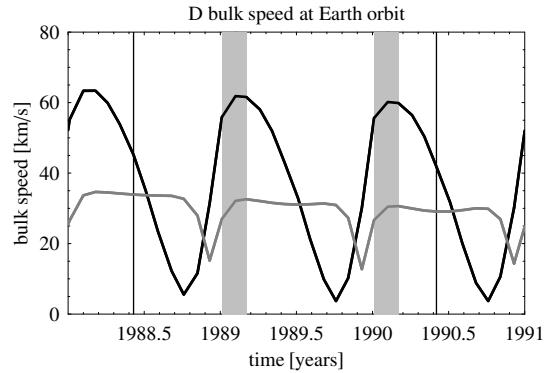} 
\caption{Bulk speed of interstellar D at Earth orbit during the phase of solar cycle corresponding to the planned operations of IBEX (gray) and relative speed of D with respect to Earth (black). Vertical lines mark the interval corresponding to IBEX observations, whereas the gray bands mark the intervals when the D beam should be visible to IBEX.}
 \label{fig:speed}
\end{figure}
\begin{figure}
\centering
\includegraphics[width=8cm]{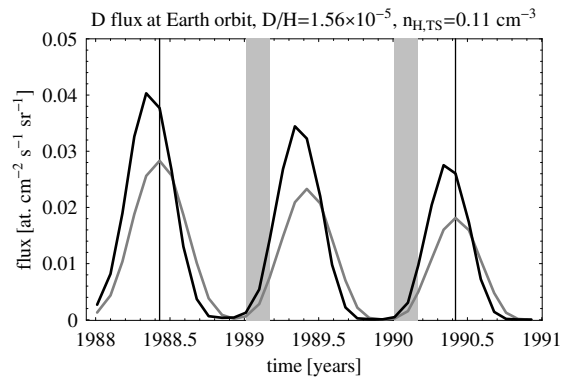} 
\caption{Net flux of neutral interstellar D expected at Earth orbit (black) and the corresponding flux relative to Earth-bound observer (gray) as function of time. Vertical lines mark the interval corresponding to the planned IBEX observations, and the gray bands mark the intervals when the D beam should be visible by IBEX.}
\label{fig:ibexMP}
\end{figure}
Our simulations indicate that deuterium abundance in Earth orbit is elevated at least twofold with respect to the abundance at TS and varies by a few orders of magnitude along the Earth orbit from upwind to downwind, where it reaches its highest value (see Fig. \ref{fig:dens}). It also varies during the solar cycle. The explanation of these reults is a strong variability of neutral interstellar hydrogen density in Earth orbit \citep{bzowski_etal:96}. The density of deuterium has a relatively mild variability (at upwind, by a factor of 0.5 from solar minimum to maximum), as shown in Fig. \ref{fig:dens}. The bulk speed of deuterium varies weakly during the solar cycle and remains almost constant along the Earth orbit (30 -- 32~km/s) apart from a tail region (gray line in Fig. \ref{fig:speed}). Consequently, the net flux of interstellar D shows an annual amplitude that is only a little larger than the amplitude of density (see the gray line in Fig.\ref{fig:ibexMP}). The amplitude decreases with the increase of solar activity, but the variabilities in the flux in the upwind and crosswind portions of Earth orbit are relatively small: the solar max/solar min flux ratios are equal to $\sim 0.4$ and $\sim 0.25$, respectively. 

The time profile of the velocity of atoms hitting a detector on an Earth-bound satellite is appreciably modified by the proper velocity of the spacecraft, which is assumed here to equal the Earth velocity, as shown by the black line in Fig. \ref{fig:speed}. Since the detection efficiency usually increases with energy, the optimal time for detection should be an interval close to the 40th day of each year, when the relative velocity is the highest. From the viewpoint of flux strength, the optimal time is about 135-th day of each year (black line in Fig. \ref{fig:ibexMP}).

\begin{figure}
\centering
\includegraphics[width=8cm]{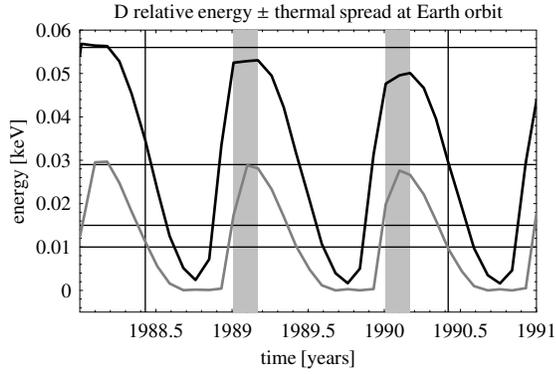} 
\caption{Relative energy of deuterium gas with respect to the moving Earth $\pm$ thermal spread. Horizontal lines mark the centers of the lowest IBEX energy channels (except the lowermost one, which marks the lower sensitivity limit), the vertical lines mark the interval corresponding to the planned IBEX observations. The ``sweet points'' for deuterium detection are marked by the gray vertical bands.}
\label{fig:ibexebf}
\end{figure}

IBEX will be a spin-stabilized Earth satellite that is designed to observe energetic neutral atoms (ENA) in the $0.01-5.9$~keV energy band, with the centers of relevant energy bands at 0.015, 0.029, 0.056, and 0.107~keV and sensors looking perpendicularly to the spin axis directed at the Sun. It is planned to operate for at least two years. The observation technique for the slowest atoms was discussed by \citet{mobius_etal:01a}. The opening angle of the IBEX-Lo instrument, best suited for observations of interstellar D, will be $7\degr$ and the rotation axis of the spacecraft will be repositioned approximately every seven days to maintain its offset from the Sun at no more than $\sim 4\degr$. Thus the beam of interstellar D atoms will be observable by IBEX when the ecliptic longitude of its vector of relative velocity with respect to Earth is inclined by $90\degr \pm 7\degr$ with respect to the Sun-Earth line. Hence, an opportunity to detect interstellar D atoms can occur potentially only twice during each year: once when IBEX is between $\sim 105\degr$ and $\sim 165\degr$ ecliptic longitude (between January and March), and the other when the spacecraft is located symmetrically with respect to the upwind longitude $254.68\degr$, i.e. between October and December. But during the latter season the relative speed between the D flow and IBEX will be so slow that the atoms will be unable to exceed the lower boundary of the energetic sensitivity band; only the January-March opportunity is therefore viable, when the Earth travels against the gas flow (Fig. \ref{fig:ibexebf}).

The beam of D atoms has a thermal spread in energy whose magnitude varies with position along the Earth orbit. The thermal spread projected onto an arbitrary line, defined by the unit vector $\vec{e}(\vec{r})$, is calculated from the formula:
\begin{equation}
	u_{T}(\vec{r}) = \left(\iiint \ \left[\left(\vec{v}(\vec{r})-\vec{u}(\vec{r})\right)\cdot\vec{e}(\vec{r})\right]^{2}\ f(\vec{r},\vec{v})\ d^{3} \vec{v}\right)^{1/2},
	\label{eq:ttens}
\end{equation}
where $\vec{u}$ is the local bulk velocity of the beam. Fig. \ref{fig:ibexebf} shows the relative energy of the D atoms incoming to IBEX within $u_{T}\left(\vec{r}\right)$ from the local velocity $u\left(\vec{r}\right)$. These atoms are located within the IBEX sensitivity limit throughout the year apart from a short interval between August and November. The beam portion contained within one thermal velocity $u_T$ from the local bulk speed $u$ will be within the IBEX sensitivity range precisely during the first half of the calendar year. The entire energy range of the beam will drop below the detectability limit of IBEX--Lo about the 230th day of each year and reemerge about DOY 320.

The lower limit to the flux detectable by the IBEX-Lo instrument is conservatively estimated to be $0.1 - 1.0$~at.~$\mathrm{cm}^{-2}\mathrm{s}^{-1}\mathrm{sr}^{-1}$ [M{\"o}bius, 2008, private comm.]. During the January--March observations opportunity, the net flux observed by IBEX is expected to increase from $0.002$ cm$^{-2}$~s$^{-1}$ to $0.015$ cm$^{-2}$~s$^{-1}$, which corresponds to the absolute fluxes equal to 0.001--0.007~cm$^{-2}$~s$^{-1}$. Under the assumptions adopted in this study, the expected flux of D is therefore at the detection threshold.

Calculations suggest that this result depends weakly on possible modifications of temperature and bulk velocity of neutral D at the termination shock. For solar minimum conditions, a change in the gas temperature at the termination shock by $\pm 6000$~K (i.e., by $\pm 27$\%) from the assumed value of $12\,000$~K produces a change in the density at 1~AU crosswind by $\pm 25$\%, and a change in bulk velocity at the termination shock of $\pm 4$~km/s (i.e., by $\pm 18$\%) from 22~km/s results in a change in the 1~AU density by $\pm 10$\%. For solar maximum conditions, these variations will be approximately twice as large when the temperature is varied, and will remain unchanged when the bulk velocity is varied. An increase in the temperature and/or bulk velocity at TS leads to an increase in deuterium density at 1~AU. On the other hand, since the local density and local flux are directly proportional to the density at the termination shock, then if the filtration of D within the heliospheric interface is weaker than in the case of hydrogen, or if the density of D in the LIC is higher than inferred from the astrophysical observations, then the density at TS might be higher and the flux at 1~AU could exceed more readily the IBEX-Lo detection threshold.

\section{Conclusions}
We performed extensive simulations of the density, velocity, and flux of neutral interstellar deuterium at Earth orbit in an attempt to check its detectability by IBEX. The conclusions from this study are the following.
\begin{enumerate}
  \item The absolute density of neutral interstellar D at Earth orbit changes during the solar cycle more weakly the corresponding H density and its abundance increases appreciably with respect to the abundance at the termination shock.
  \item The bulk velocity of D at Earth orbit depends weakly on the ecliptic longitude (apart from a narrow cone in the downwind region), so variations of the absolute flux follow the variations of density.
 \item The relative energy of D, and its thermal spread, are within the IBEX sensitivity band apart from an interval between August and November each year, but because of the measurement geometry the only observation season during the year occurs when the Earth is traveling ``against the wind" of interstellar D at Earth orbit, i.e. between January and March.
  \item The expected relative flux is equal to $0.015$~cm$^{-2}$~s$^{-1}$, which corresponds to the absolute flux equal to $0.007$~cm$^{-2}$~s$^{-1}$, assuming the H density at TS equal to 0.11~cm$^{-3}$ and a D/H abundance in the gas phase at TS equal to 15.6 ppm.
 \item Under these assumptions, the expected flux of D appears to be at the sensitivity limit of the IBEX-Lo instrument. 
\end{enumerate}

\begin{acknowledgements}
We gratefully acknowledge the encouragement from Eberhard M{\"o}bius to carry out this study and the insightful and informative discussions on the IBEX-Lo instrument. We are indebted to Andrzej So{\l}tan for pointing us an important sign mistake in the calculations. This work was supported by  Polish grants 1P03D00927 and N522 002 31/0902. M.B. gratefully acknowledges the hospitality of ISSI in Bern, where a part of this work was carried out within the framework of a Working Group on Heliospheric Breathing.
\end{acknowledgements}

\bibliographystyle{aa}
\bibliography{iplbib}

\begin{thebibliography}{24}
\expandafter\ifx\csname natexlab\endcsname\relax\def\natexlab#1{#1}\fi

\bibitem[{Bzowski(2001)}]{bzowski:01a}
Bzowski, M. 2001, in The {O}uter {H}eliosphere: {T}he {N}ext {F}rontiers, ed.
  K.~Scherer, H.~Fichtner, H.~J. Fahr, \& E.~Marsch, COSPAR Coll. Ser.
  Vol. 11,  69--72 

\bibitem[{Bzowski {et~al.}(1996)Bzowski, Fahr, \&
  Ruci{\'n}ski}]{bzowski_etal:96}
Bzowski, M., Fahr, H.~J., \& Ruci{\'n}ski, D. 1996, Icarus, 124, 209

\bibitem[{Bzowski {et~al.}(1997)Bzowski, Fahr, Ruci{\'n}ski, \&
  Scherer}]{bzowski_etal:97}
Bzowski, M., Fahr, H.~J., Ruci{\'n}ski, D., \& Scherer, H. 1997, \aap, 326, 396 
\bibitem[{{Bzowski} {et~al.}(2008){Bzowski}, {M{\"o}bius}, {Tarnopolski},
  {Izmodenov}, \& {Gloeckler}}]{bzowski_etal:08a}
{Bzowski}, M., {M{\"o}bius}, E., {Tarnopolski}, S., {Izmodenov}, V., \&
  {Gloeckler}, G. 2008, ArXiv e-prints, 0710.1480 (submitted to A\&A)

\bibitem[{Costa {et~al.}(1999)Costa, Lallement, Qu{\'e}merais, Bertaux,
  Kyr{\"o}l{\"a}, \& Schmidt}]{costa_etal:99}
Costa, J., Lallement, R., Qu{\'e}merais, E., {et~al.} 1999, \aap, 349, 660

\bibitem[{Fahr(1978)}]{fahr:78}
Fahr, H.~J. 1978, \aap, 66, 103

\bibitem[{Fahr(1979)}]{fahr:79}
Fahr, H.~J. 1979, \aap, 77, 101

\bibitem[{Fahr {et~al.}(1995)Fahr, Osterbart, \& Ruci{\'n}ski}]{fahr_etal:95}
Fahr, H.~J., Osterbart, R., \& Ruci{\'n}ski, D. 1995, \aap, 294, 587

\bibitem[{Izmodenov(2001)}]{izmodenov:01}
Izmodenov, V.~V. 2001, \ssr, 97, 385

\bibitem[{Izmodenov {et~al.}(1997)Izmodenov, Lallement, \&
  Malama}]{izmodenov_etal:97a}
Izmodenov, V.~V., Lallement, R., \& Malama, Y.~G. 1997, \aap, 317, 193

\bibitem[{Lemaire {et~al.}(2002)Lemaire, Emerich, Vial, Curdt, Sch{\"u}le, \&
  Wilhelm}]{lemaire_etal:02}
Lemaire, P.~L., Emerich, C., Vial, J.~C., {et~al.} 2002, in ESA SP-508, 219--222

\bibitem[{Linsky(2007)}]{linsky:07}
Linsky, J.~L. 2007, \ssr, 130, 367

\bibitem[{{Linsky} {et~al.}(2006){Linsky}, {Draine}, {Moos}, {Jenkins}, {Wood},
  {Oliveira}, {Blair}, {Friedman}, {Gry}, {Knauth}, {Kruk}, {Lacour}, {Lehner},
  {Redfield}, {Shull}, {Sonneborn}, \& {Williger}}]{linsky_etal:06a}
{Linsky}, J.~L., {Draine}, B.~T., {Moos}, H.~W., {et~al.} 2006, \apj, 647, 1106

\bibitem[{{McComas} {et~al.}(2004){McComas}, {Allegrini}, {Bartolone},
  {Bochsler}, {Bzowski}, {Collier}, {Fahr}, {Fichtner}, {Frisch}, {Funsten},
  {Fuselier}, {Gloeckler}, {Gruntman}, {Izmodenov}, {Knappenberger}, {Lee},
  {Livi}, {Mitchell}, {M{\" o}bius}, {Moore}, {Reisenfeld}, {Roelof},
  {Schwadron}, {Wieser}, {Witte}, {Wurz}, \& {Zank}}]{mccomas_etal:04a}
{McComas}, D., {Allegrini}, F., {Bartolone}, L., {et~al.} 2004, in AIP Conf. Ser., Vol. 719, 162--181

\bibitem[{{McComas} {et~al.}(2005){McComas}, {Allegrini}, {Bartolone},
  {Bochsler}, {Bzowski}, {Collier}, {Fahr}, {Fichtner}, {Frisch}, {Funsten},
  {Fuselier}, {Gloeckler}, {Gruntman}, {Izmodenov}, {Knappenberger}, {Lee},
  {Livi}, {Mitchell}, {M{\" o}bius}, {Moore}, {Pope}, {Reisenfeld}, {Roelof},
  {Runge}, {Scherrer}, {Schwadron}, {Tyler}, {Wieser}, {Witte}, {Wurz}, \&
  {Zank}}]{mccomas_etal:05a}
{McComas}, D., {Allegrini}, F., {Bartolone}, L., {et~al.} 2005, in ESA-SP,
  Vol. 592, Solar Wind 11/SOHO 16, 
  689--692

\bibitem[{{McComas} {et~al.}(2006){McComas}, {Allegrini}, {Bartolone},
  {Bochsler}, {Bzowski}, {Collier}, {Fahr}, {Fichtner}, {Frisch}, {Funsten},
  {Fuselier}, {Gloeckler}, {Gruntman}, {Izmodenov}, {Knappenberger}, {Lee},
  {Livi}, {Mitchell}, {M{\"o}bius}, {Moore}, {Pope}, {Reisenfeld}, {Roelof},
  {Runge}, {Scherrer}, {Schwadron}, {Tyler}, {Wieser}, {Witte}, {Wurz}, \&
  {Zank}}]{mccomas_etal:06}
{McComas}, D.~J., {Allegrini}, F., {Bartolone}, L., {et~al.} 2006, in AIP
  Conf. Ser., Vol. 858, 241--250 \

\bibitem[{M{\"o}bius {et~al.}(2001)M{\"o}bius, Litvinenko, Saul, Bzowski, \&
  Ruci{\'n}ski}]{mobius_etal:01a}
M{\"o}bius, E., Litvinenko, Y., Saul, L., Bzowski, M., \& Ruci{\'n}ski, D.
  2001, COSPAR Coll. Ser., Vol. 11, 69--72

\bibitem[{Osterbart \& Fahr(1992)}]{osterbart_fahr:92}
Osterbart, R. \& Fahr, H.~J. 1992, \aap, 264, 260

\bibitem[{Ruci{\'n}ski \& Bzowski(1995)}]{rucinski_bzowski:95b}
Ruci{\'n}ski, D. \& Bzowski, M. 1995, \aap, 296, 248

\bibitem[{{Tarnopolski} \& {Bzowski}(2007)}]{tarnopolski_bzowski:07a}
{Tarnopolski}, S. \& {Bzowski}, M. 2007, ArXiv Astrophysics e-prints 0701133 (submitted to A\&A)

\bibitem[{{Tarnopolski}(2007)}]{tarnopolski:07}
{Tarnopolski}, S.~T. 2007, PhD thesis, {Space Research Centre PAS}

\bibitem[{Thomas(1978)}]{thomas:78}
Thomas, G.~E. 1978, Ann. Rev. Earth Planet. Sci., 6, 173

\bibitem[{{Tobiska} {et~al.}(2000){Tobiska}, {Woods}, {Eparvier}, {Viereck},
  {Floyd}, {Bouwer}, {Rottman}, \& {White}}]{tobiska_etal:00c}
{Tobiska}, W.~K., {Woods}, T., {Eparvier}, F., {et~al.} 2000, J. Atm. Terr.
  Phys., 62, 1233

\bibitem[{{Witte}(2004)}]{witte:04}
{Witte}, M. 2004, \aap, 426, 835

\end{thebibliography}

\end{document}